\documentclass[3p,onecolumn,floatfix]{elsarticle}
\usepackage{lineno,hyperref}
\usepackage{graphicx}
\usepackage{url}
\usepackage{amsmath}
\usepackage{xcolor}
\usepackage[normalem]{ulem}
\modulolinenumbers[5]
\usepackage[T1]{fontenc}

\journal{Nuclear Instruments and Methods in Physics Research A}

\begin{document}

\begin{frontmatter}

\title{Feasibility studies for imaging $e^{+}e^{-}$ annihilation with modular multi-strip detectors}

\author[IOP,TBP,CFT]{S.~Sharma\corref{cor1}}
\ead{sushil.sharma@uj.edu.pl}
\cortext[cor1]{sushil.sharma@uj.edu.pl}
\author[UOT,TIF]{L.~Povolo}
\author[UOT,TIF]{S.~Mariazzi}
\author[IOP,TBP]{G.~Korcyl}
\author[IOP,TBP,CFT]{K.~Kacprzak}
\author[IOP,TBP,CFT]{D.~Kumar}
\author[IOP,TBP,CFT]{S.~Nied{\'z}wiecki}
\author[IOP,TBP,CFT]{J.~Baran}
\author[IOP,TBP,CFT]{E.~Beyene}
\author[UOT,TIF]{R.S.~Brusa}
\author[TIF]{R.~Caravita}
\author[IOP,TBP,CFT]{N.~Chug}
\author[IOP,TBP,CFT]{A.~Coussat}
\author[INF]{C.~Curceanu}
\author[IOP,TBP,CFT]{E.~Czerwinski}
\author[IOP,TBP,CFT]{M.~Dadgar}
\author[IOP,TBP,CFT]{M.~Das}
\author[IOP,TBP,CFT,INF]{K.~Dulski}
\author[IOP,TBP,CFT]{K.~Eliyan}
\author[IOP,TBP,CFT]{A.~Gajos}
\author[IOP,TBP,CFT]{N.~Gupta}
\author[UOV]{B.C.~Hiesmayr}
\author[IOP,TBP,CFT]{{\L}.~Kap{\l}on}
\author[IOP,TBP,CFT]{T.~Kaplanoglu}
\author[DCS]{K.~Klimaszewski}
\author[DCS]{P.~Konieczka}
\author[IOP]{T.~Kozik}
\author[INP]{M.K.~Kozani}
\author[HEP]{W.~Krzemie{\'n}}
\author[IOP,TBP,CFT]{S.~Moyo}
\author[IOP,TBP,CFT]{W.~Mryka}
\author[UOT,TIF]{L.~Penasa}
\author[IOP,TBP,CFT]{S.~Parzych}
\author[IOP,TBP,CFT]{E.~Perez.~Del Rio}
\author[DCS]{L.~Raczy\'nski}
\author[IOP,TBP,CFT]{Shivani}
\author[DCS]{R.Y~Shopa}
\author[IOP,TBP,CFT]{M.~Skurzok}
\author[IOP,TBP,CFT]{E.{\L}.~St{\k{e}}pie\'n}
\author[IOP,TBP,CFT]{P.~Tanty}
\author[IOP,TBP,CFT]{F.~Tayefi}
\author[IOP,TBP,CFT]{K.~Tayefi}
\author[HEP]{W.~Wi{\'s}licki}
\author[IOP,TBP,CFT]{P.~Moskal}

\address[IOP]{Faculty of Physics, Astronomy and Applied Computer Science, Jagiellonian University, 30-348 Krak\'ow, Poland}
\address[TBP]{Total-Body Jagiellonian-PET Laboratory, Jagiellonian University, Krak\'ow, Poland}
\address[CFT]{Center for Theranostics, Jagiellonian University, Krak\'ow, Poland}
\address[UOT]{Department of Physics, University of Trento, via Sommarive 14, 38123 Povo, Trento, Italy}
\address[TIF]{TIFPA/INFN, via Sommarive 14, 38123 Povo, Trento, Italy}
\address[INF]{INFN, Laboratori Nazionali di Frascati CP 13,  Via E. Fermi 40, 00044, Frascati, Italy}
\address[UOV]{Faculty of Physics, University of Vienna, W$\ddot{a}$hringerstrasse 17, 1090$-$Vienna, Vienna, Austria}
\address[DCS]{Department of Complex Systems, National Centre for Nuclear Research, 05-400 Otwock-Świerk, Poland}
\address[HEP]{High Energy Physics Division, National Centre for Nuclear Research, 05-400 Otwock-Świerk, Poland}
\address[INP]{Institute of Nuclear Physics, Polish Academy of Sciences, 31-342 Krak\'ow, Poland}


\begin{abstract}
Studies based on imaging the annihilation of the electron ($e^{-}$) and its antiparticle positron ($e^{+}$) open up several interesting applications in nuclear medicine and fundamental research. The annihilation process involves both the direct conversion of $e^{+}e^{-}$ into photons and the formation of their atomically bound state, the positronium atom (Ps), which can be used as a probe for fundamental studies.~With the ability to produce large quantities of Ps, manipulate them in a long-lived Ps states, and image their annihilations after a free fall or after passing through atomic interferometers, this purely leptonic antimatter system can be used to perform inertial sensing studies in view of a direct test of Einstein's equivalence principle. It is envisioned that modular multi-strip detectors can be exploited as potential detection units for this kind of studies. In this work, we report the results of the first feasibility study performed on a $e^{+}$ beamline using two detection modules to evaluate their reconstruction performance and spatial resolution for imaging $e^{+}e^{-}$ annihilations and thus their applicability for gravitational studies of Ps.
\end{abstract}
\end{frontmatter}

\section{Introduction}
The positron ($e^{+}$) is the lightest stable antiparticle and differs from other antimatter objects primarily in the sense that other antimatter objects require an accelerator for their creation in the laboratory~\cite{Charpak1960,Baur1993}. Since it is relatively easy to obtain $e^{+}$ either by pair production processes or in $\beta^{+}$ radioactive decays, it became popular shortly after its discovery~\cite{Anderson1932}. High-energy positrons are produced on a large scale in accelerator facilities such as the Large Hadron Collider (LHC)~\cite{Arikan} and the Beijing Electron Positron Collider Upgrade (BEPCII)~\cite{Wang}. The study of their collisions with electrons or protons enables the exploration of the fundamental constituents of matter, the study of particle interactions, and even the search for new particles or phenomena beyond the Standard Model (BSM). While low-energy positrons (up to tens of keV) have a variety of applications, such as a non-invasive tracer in medical imaging~\cite{Chery2004, Moskal21, Shopa22, Alavi2021, Moskal2019}, $e^{+}$ annihilation-based techniques are used to identify and characterise defects in materials~\cite{Mijnarends}, and in fundamental physics~\cite{Moskal16, Moskal21B}. Moreover, $e^{+}$  can form a metastable atom when interacting with $e^{-}$, the positronium atom (Ps)~\cite{Deutsch1953}, which is a purely leptonic object and an excellent two-body system for testing non-relativistic quantum electrodynamics (nrQED) in the bound state~\cite{Cassidy2018,Bass2019}. Ps can be formed in one of two possible ground states: spin 0 state, known as para-Ps (p-Ps, $^{1}S_{0}$), which is short-lived (125~ps), or spin 1 state, long-lived state of Ps (142~ns), also known as ortho-Ps (o-Ps, $^{3}S_{1}$)~\cite{Deutsch1953}. The study of the decays of ortho-positronium atoms has been used for a deeper understanding of the fundamental symmetries~\cite{Moskal21B,Bass2023}.\newline
With the ability to populate o-Ps in excited states through laser manipulation~\cite{Cassidy2018,Deller2016}, its lifetime can be enhanced by more than one order of magnitude~\cite{Alonso2017,Amsler}. Positronium atoms in Rydberg or 2$^{3}$S Ps state have been postulated as a potential probe for performing inertial sensing studies towards a direct test of Einstein's equivalence principle on antimatter~\cite{Mills2002,Cassidy2014,Oberthaler}. In particular, the proposed studies on 2$^{3}$S Ps are based on the application of the technique of atomic interferometry/deflectometry to measure gravitational effects on Ps atoms. The experimental scheme described by Mariazzi et al.~\cite{Mariazzi} requires a beam of 2$^{3}$S Ps atoms, optimization of the parameters for the interferometer setup, and position-sensitive detectors with sub-nm spatial resolution to study the fringe pattern formed as the Ps atoms pass through the interferometers. As suggested in Ref.~\cite{Mariazzi}, such resolution could be achieved by scanning the fringe pattern with a material grating of the same periodicity of the fringe moved by a piezoelectric actuator. A position sensitive detector can be used to count the Ps annihilations on the grating~\cite{Mariazzi22}. An additional stopper can be placed 10-20~mm behind the moving grating to detect the Ps atoms crossing the grating. Ps annihilations on the grating and on the stopper can be distinguished if the spatial resolution of the detector is better than the distance between the grating and stopper. This requirement places a limitation on the detector to be used, as it should have a spatial resolution of 10-20 mm. Imaging techniques must be used to reconstruct the annihilation vertices. Therefore, detectors with good time-of-flight (TOF) resolution are preferable. The modular multi-strip detection units based on plastic scintillators developed by the J-PET (Jagiellonian-PET) collaboration~\cite{Moskal14, Szymon17, Moskal2021Bio, Sharma2023B, Kaplon} are a good solution for this type of measurements. The detection modules can be operated individually or in pairs and are suitable as position-sensitive detectors for inertial sensing studies to reconstruct annihilation vertices. To investigate the feasibility of imaging $e^{+}e^{-}$ annihilations with two modular detection units and to evaluate their reconstruction performance, a pilot measurement was performed at the $e^{+}$ beamline of the Anti-Matter Laboratory (AML) in Trento. The characteristic details of the detectors will be discussed in the next section. Section 3 describes the experimental details, followed by the results (in Section 4). Section 5 provides the summary and an outlook of the studies.  
%
\section{Modular Jagiellonian Positron Emission Tomograph (Modular J-PET)}
\begin{figure}[!ht]
\centering
\includegraphics[scale=.44]{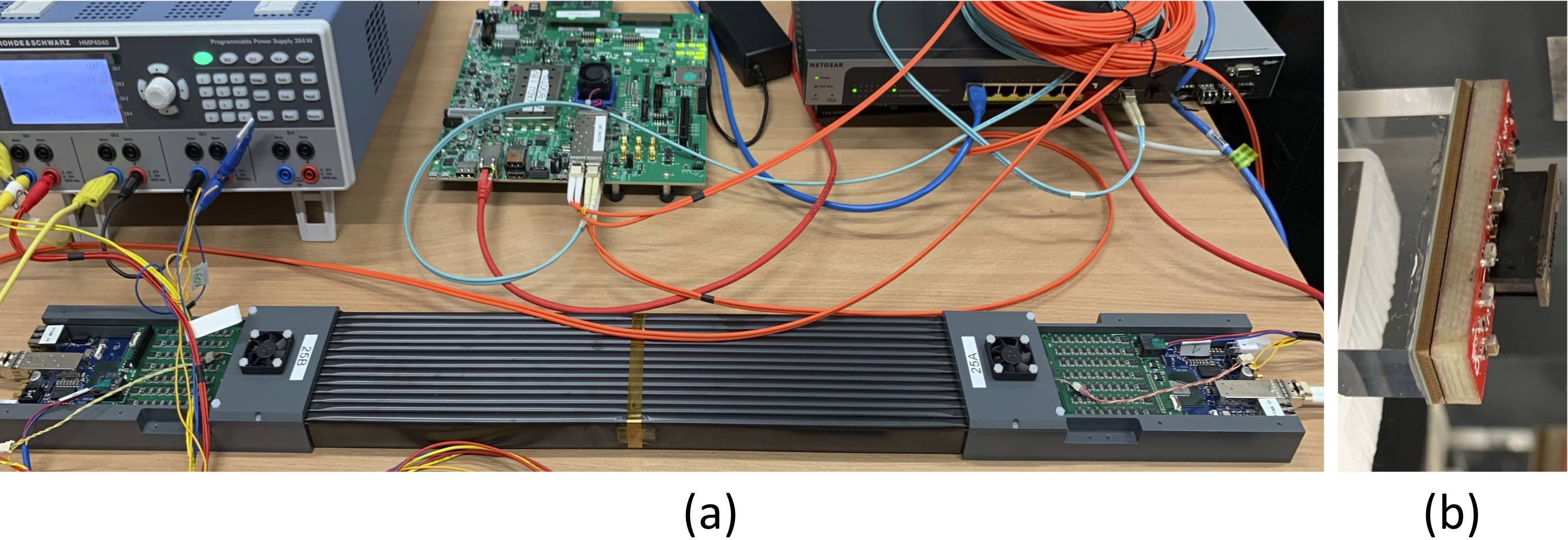}
\caption{(a) Shows a single module with complete signal readout chain. The individual scintillators are wrapped first with Vikuiti and then with black foils. (b) represents the edge of the scintillator before the wrapping, which is attached to the matrix of SiPMs.}
\label{modular}
\end{figure}

The modular J-PET is based on the design of stand-alone detection modules with connected front-end electronics (see Fig.~\ref{modular} (a)). Each module consists of densely packed 13 plastic scintillators of dimension 500$\times$24$\times$6~mm$^{3}$ glued on both sides to a 1$\times$4 matrix of silicon photomultipliers (SiPMs) (Fig.~\ref{modular}(a,b)).    
The signals from the SiPMs are read out using a newly developed electronic front-end board that enables signal sampling in the voltage domain with an accuracy of 20~ps \cite{Palka}. Data are stored with Field Programmable Gate Array (FPGAs) in triggerless mode, which is easily reconfigurable~\cite{Korcyl}. For the estimation of energy deposition by the photon interaction inside plastic scintillators, the time-over-threshold method (TOT) is adapted instead of the charge collection method~\cite{Sharma20}. The hit position and hit time are estimated by measuring the arrival time of light signals at each end of the scintillator~\cite{Moskal14}. The total length of a single module is 90.6~cm, including the length of the associated front-end electronic boards, and the width is 9~cm. A single module weighs less than 2~kg. The J-PET collaboration constructed 24 of such detection modules for positron emission tomography applications, which can be assembled into a cylinder with a diameter of 76.2~cm and an axial field of view of 50~cm~\cite{Moskal20B}. The applications of detection modules with positron and positronium beams have been discussed in a previous work \cite{Sharma}. The technical details of the modules and the algorithm for data analysis are presented in the section~\ref{resultspart}.
%
\section{Experimental setup and data measurement}
To evaluate the performance of modular detectors for the reconstruction of the vertices of $e^{+}$ annihilations, two detection units were brought to the AML laboratory of the University of Trento in Italy. A new beamline has recently been commissioned that can deliver a continuous positron beam with a spot diameter of less than 5~mm. The details of the positron beamline will be reported elsewhere~\cite{Luca}.  
\begin{figure}[!ht]
\centering
\includegraphics[scale=.48]{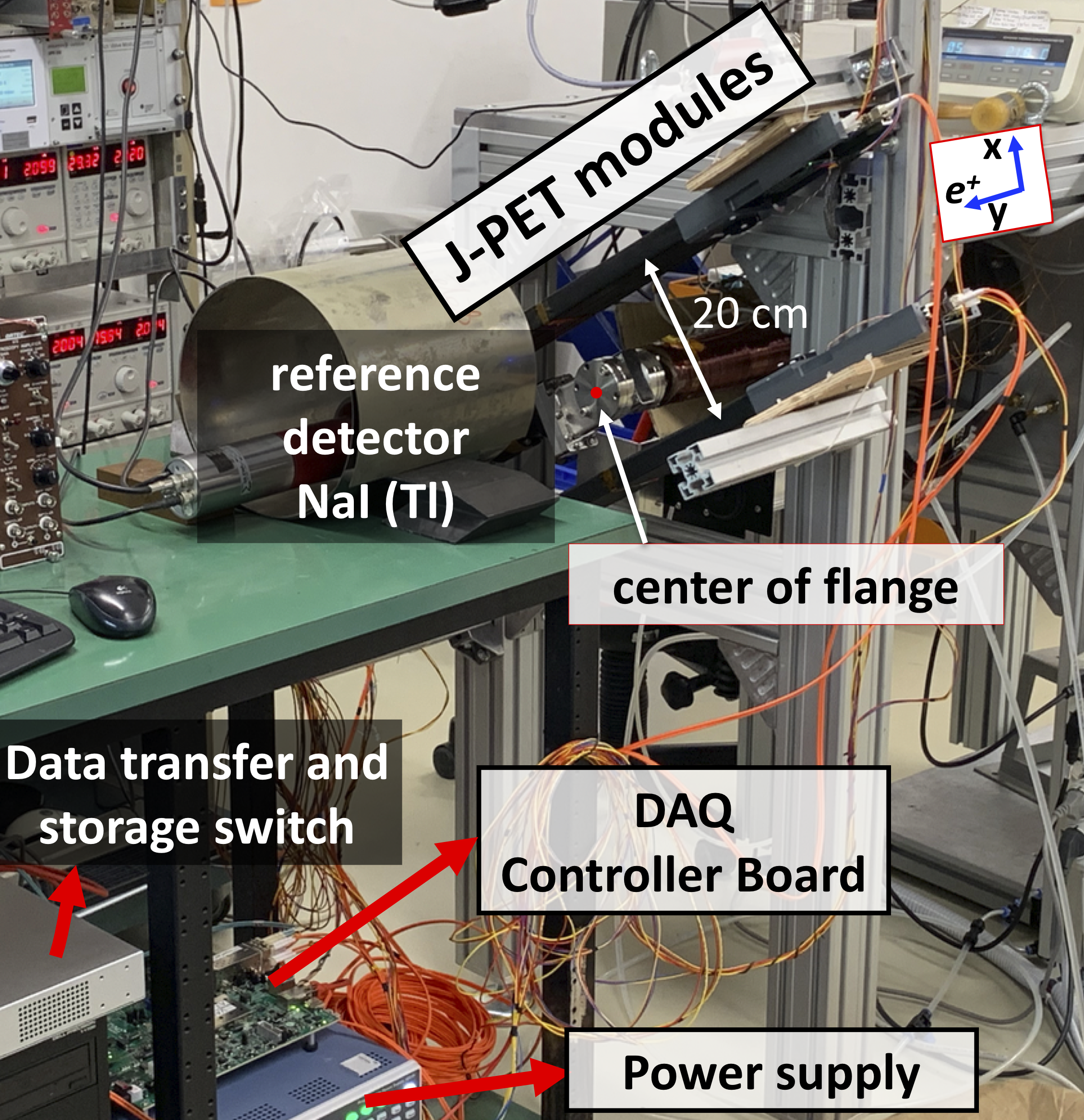}
\caption{The picture shows the experimental setup with two modular J-PET detection modules placed 10 cm apart on either side of the positron beam and centered on a flange where positrons are to be annihilated (red dot on the flange). A NaI(Tl) single crystal  detector is placed at 25~cm distance from the flange, aligned with the axis of the $e^{+}$ beam. Data are acquisited by an FPGA-based controller board and eventually stored on the computer hard disk via a fast data transfer switch.}
\label{experimentalsetup}
\end{figure}
To perform the experiment, a flange was used as a beam terminator, representing the origin surface of the two counter-propagating 511~keV photons from $e^{+}$ annihilation. The annihilation photons were registered by modular J-PET detection units placed on each side at a distance of 10~cm from the centre of the flange where the annihilation spots are expected to be (Fig.~\ref{experimentalsetup}). The red dot in the image shows the centre of the flange. The signals from the SiPMs are processed by FTAB boards (combination of front-end electronics, TDCs and readouts channels) using an FPGA-based controller board and stored in external memory via a fast data transfer switch. In addition, to monitor the $e^{+}$ beam rate, a NaI(Tl) single crystal of dimension 3$^{"}\times3^{"}$ was aligned at a distance of about 25~cm behind the flange. The crystal was surrounded by a 5~mm thick cylindrical tungsten shield to reduce the number of unwanted counts. Positron rate was recorded every 10 minutes by integrating the 511 keV photon peak after correction for background and attenuation factor caused by the flange material. 
%
\section{Results}\label{resultspart} 

\subsection{Low-level data reconstruction}
The binary data recorded by the FPGA boards are processed using the dedicated data analysis framework developed by the J-PET collaboration~\cite{Krzemien}. The procedures are divided into steps that start with reading the timestamps of the DAQ channels and end with categorized physical events for the further analysis. The data were collected in triggerless mode and the binary data packets are collected in time slots of 50~microseconds. 
\begin{figure}[!ht]
\centering
\includegraphics[scale=.49]{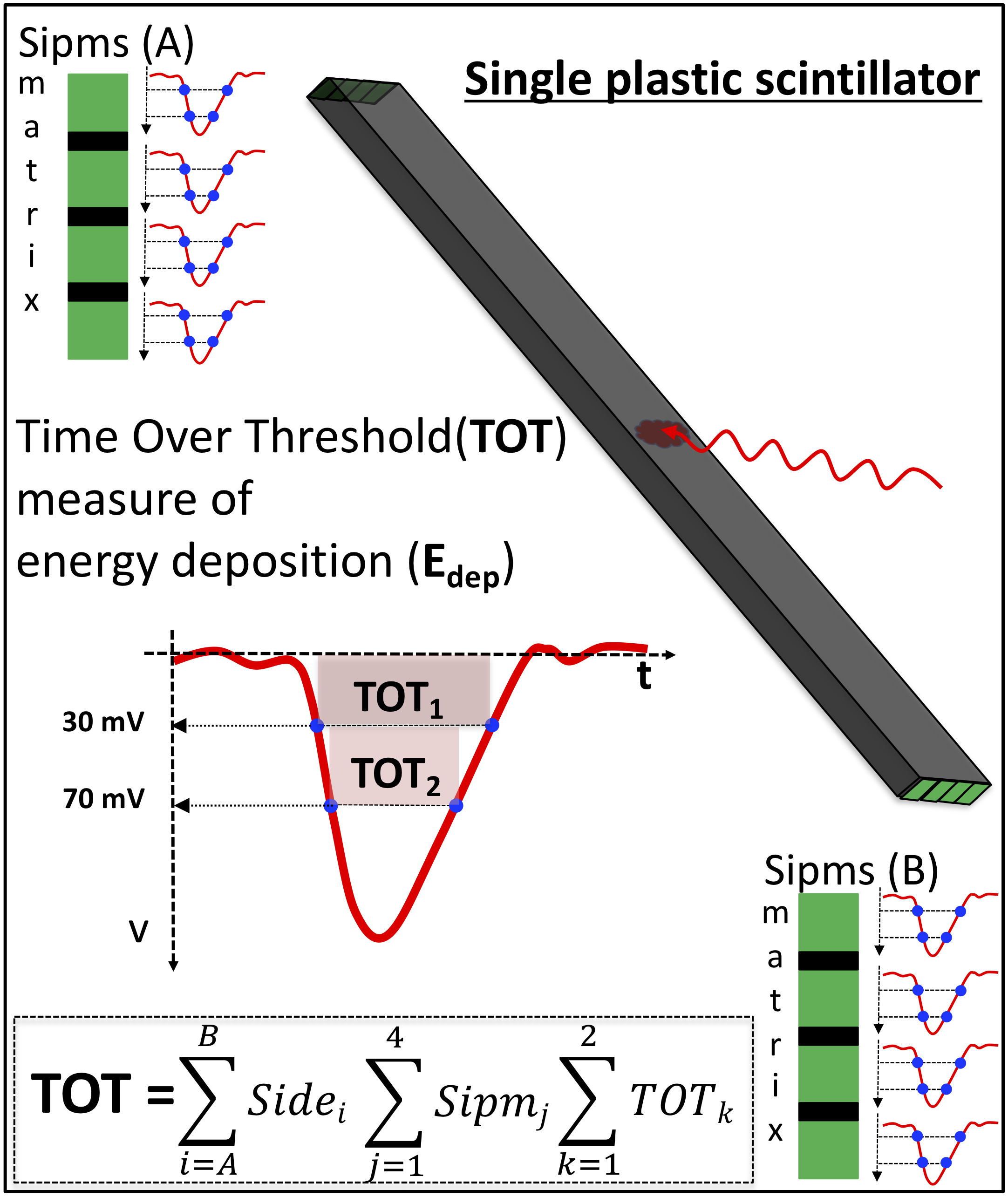}
\caption{Shows the schematic of signal readout through a 1x4 matrix of SiPMs of a plastic scintillator at both edges. A~photon interacts with the scintillator and deposits a certain energy. Based on the deposited energy, the signals obtained at each SiPM are probed at two thresholds of 30~mV and 70~mV. The deposited energy can be estimated by summing the TOTs calculated for the signals from all SiPMs at the fixed thresholds on either side of the scintillators. A signal from a SiPM is only considered if it has an amplitude higher than the lowest applied threshold (30~mV). The formula in the figure shows the case where the signal amplitude of all SiPMs on both sides of the scintillator exceeds the two applied thresholds for the calculation of TOT, which is used as a measure of energy deposition(Edep).}
\label{single}
\end{figure}
The signals are reconstructed for each SiPM using the measured timestamps at two thresholds (30~mV and 70~mV) for rising and falling edges, TOT is calculated using a rectangular approximation, as shown in Fig.~\ref{single}. Signals from up to 4 SiPMs located at the end of each strip are combined into a matrix signal. The arrival time of the signal is calculated using the average values of the SiPMs signals found within 1.3~ns coincidence. The average of the measured TOT values on all SiPMs on each side of the scintillators gives the measure of energy deposition for a given interaction. 
For the $e^{+}$ annihilation spot imaging we are using line-of-response (LOR) and time-of-flight (TOF), which require the reconstruction of time and position of photon interaction in scintillating strips. For J-PET modules, these two observables are estimated based on the measured time difference of the light signals arriving at both ends of the scintillators and read out by photomultipliers, and the estimated value of effective velocity of light in plastic strips \cite{Moskal14}.
\subsection{Calibration of the detector}
\begin{figure}[!hb]
\centering
\includegraphics[scale=.58]{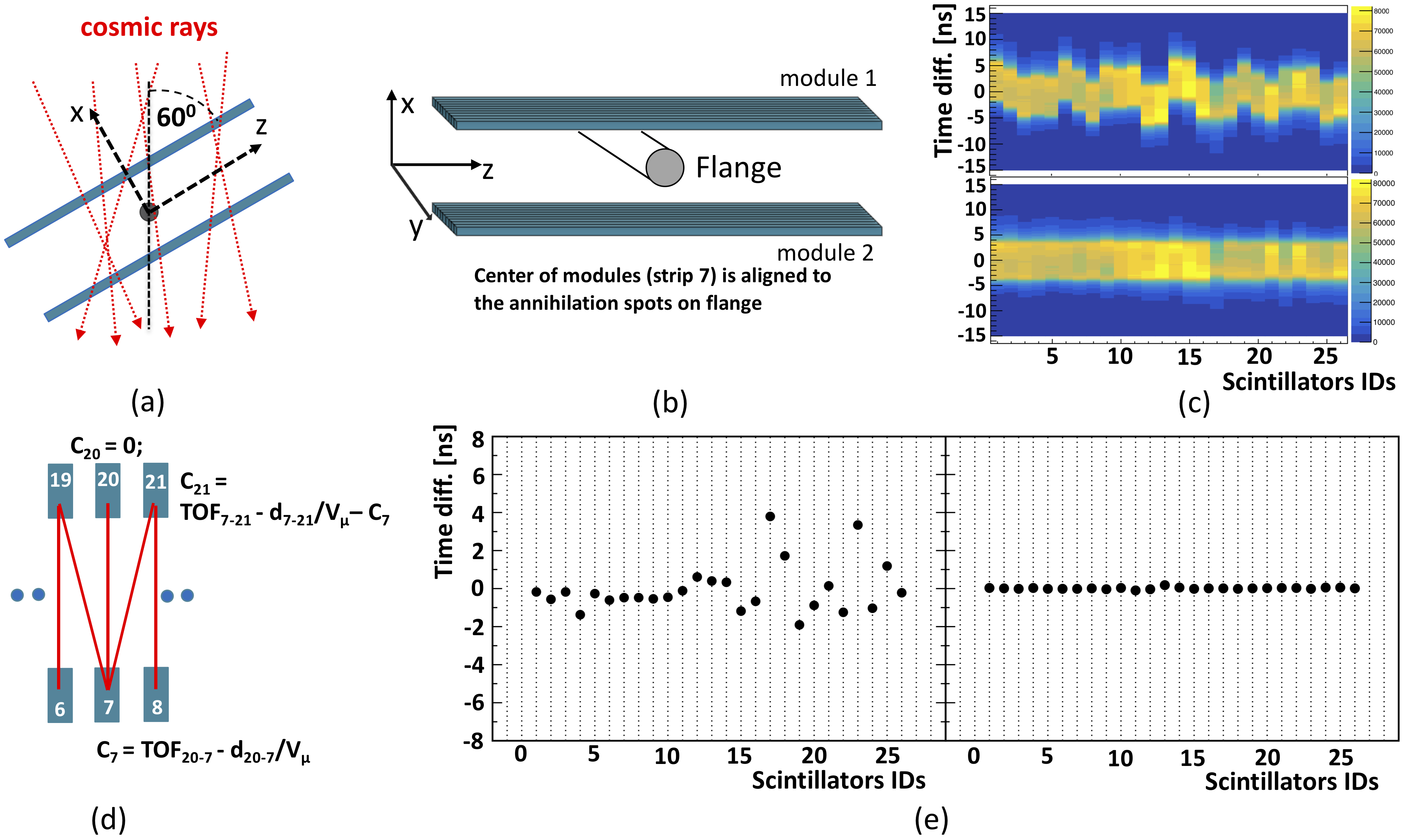}
\caption{Calibration procedure used to synchronise the timing information as well as the TOF of the plastic scintialltors. (a) Shows the placement of two detection modules around the $e^{+}$ beam, with the red dashed lines representing the cosmic showers on the modules. (b) Coordinate system used for calibration and later for data analysis. The results of the hit time differences of the individual signals before (upper panel) and after (lower panel) the calibrations are shown in (c). In addition, the TOF offsets for the scintillators of the modules were calibrated according to the scheme described in (d). Values of the corrections for the TOF offsets for each scintillator strip are shown in (e), first iteration (left panel) and 20$^{th}$ iteration (right panel) calibration. The details of the calibration method are explained in the text.}
\label{calibration}
\end{figure}

The calibration of the electronic offsets for estimation of interaction time in plastic scintillators was performed using cosmic rays, when the e$^{+}$ beam was off. Figure~\ref{calibration}(a) shows the pictorial representation of the detector placement in the experimental setup, where cosmic rays irradiate all scintillators equally. Due to the experimental constraint, the detectors were placed with a vertical angular displacement of 60$^{\circ}$. For the analysis of the measured data, the coordinate system was rearranged as right-handed contention~(Fig.~\ref{calibration}(b)), fixing the x-axis in the upward direction, the y-axis in the direction of the beam, and the z-axis along the axial length of the scintillators. 
%
%
%
The first step was to synchronise the time differences between signals registered at the opposite ends of the strips. To do this, the slopes of the time difference signal edges for each strip were compared and aligned. The top panel in Fig.~\ref{calibration} (c) shows the measured hit time difference of the signals from the scintillators versus their IDs before calibration and the panel below after calibration. The scintillators in both modules were assigned unique IDs ranging from 1 to 26 (each module has 13 scintillators). The calibration of TOF offsets for each strip is based on the selection of a pair of muon hits, with one hit occurring in the plastic scintillator of the upper module and the other in the lower. Assuming average velocity ($V_\mu$) of the muons is 29.8 cm/ns~\cite{LULU}, we could calculate the TOF offsets for vertically aligned or adjacent strips according to the scheme shown in Fig.~\ref{calibration} (d). The offset for the middle strip (e.g. with ID = 20 of the upper module) was set to 0. Then the offset of the middle strip of the lower module (C$_{7}$) can be estimated by calculating the difference between the measured TOF and the estimated TOF (d$_{20-7}$/$V_\mu$), where d$_{20-7}$ is the distance that muon travels when interacting with strips IDs 20 and 7. For the neighbouring plastic strips, the estimated TOF offset was corrected, as shown in the example for strip ID 21 (C$_{21}$). The same scheme was chosen to calculate the offsets of the other scintillators. The estimated offsets were optimised using the iterative approach. The procedure was repeated until the corrections calculated in the iteration were smaller than 50~ps. Figure~\ref{calibration}(e) shows these corrections to TOF offsets as a function of scintillator IDs for first (left panel) and 20$^{th}$ iteration (right panel).   
\subsection{Data analysis for Imaging $e^{+}e^{-}$ annihilations}
\begin{figure}[!ht]
\centering
\includegraphics[scale=.70]{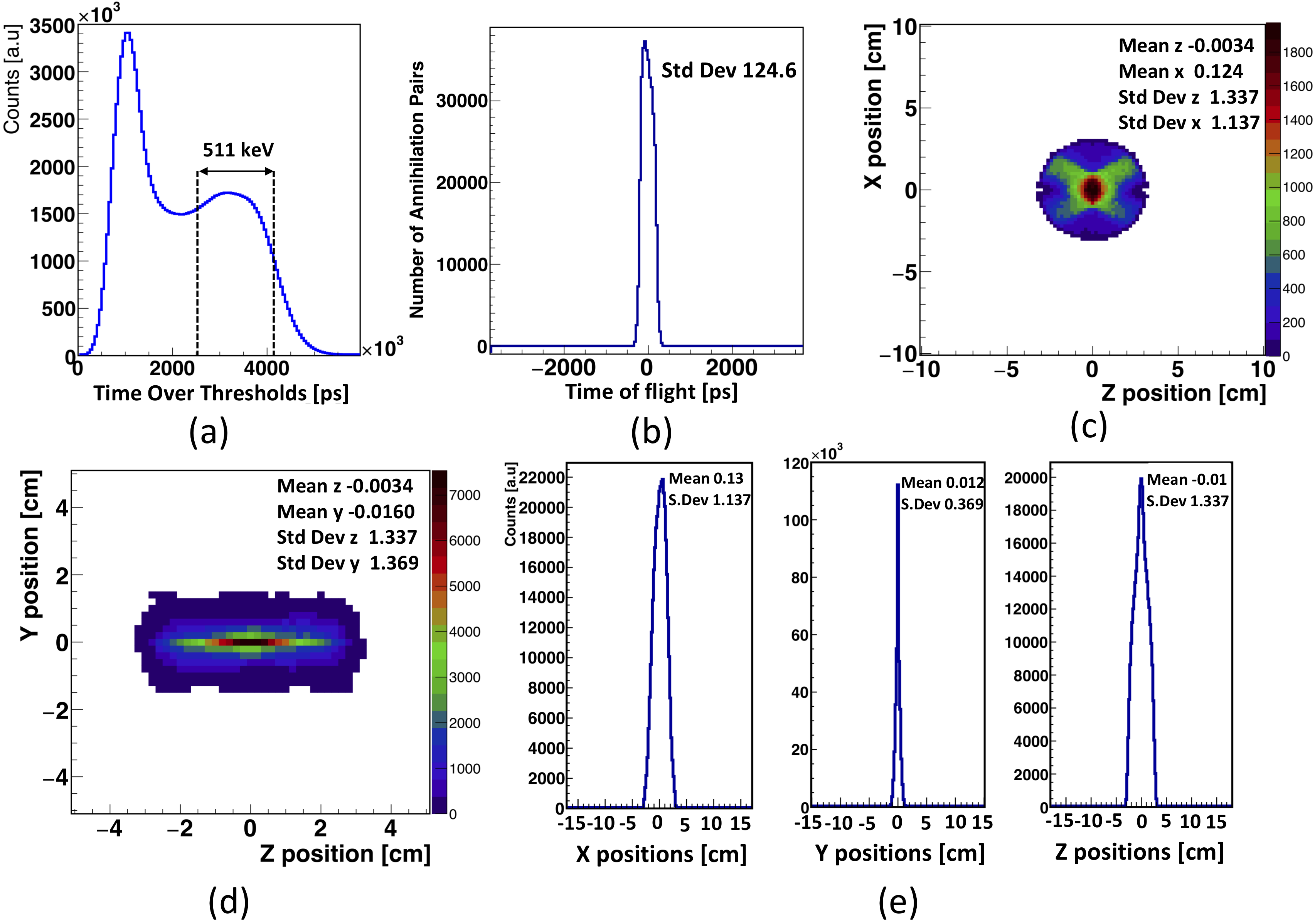}
\caption{(a) Shows the TOT spectrum, which is a measure of energy deposition. Since the photons interact with the plastic scintillator mainly via Compton scattering, a Compton edge is expected for the maximum energy deposition, which is visible in the figure. The peak at lower TOT values corresponds to the lower energy deposits. Selecting the window of TOT values around the Compton edge, as shown in the figure, allows the interactions to be marked by 511 keV photons. (b) TOF spectrum estimated for the selected annihilation pairs. (c,d) show the images of the annihilation spots in the zx and zy planes reconstructed by the developed algorithm, and (e) show the projections of the reconstructed images in the x, y, and z axes, respectively.} 
\label{results}
\end{figure}

For the reconstruction of $e^{+}$ annihilation vertices, an algorithm has been  developed to analyse events with 2-hits expected from 511 keV photons. The first selection criterion for choosing annihilation photons is based on the energy deposition measured as TOT in the context of J-PET data analysis framework. Fig.~\ref{results}(a) shows a typical TOT spectra obtained for 511~keV photons. Hits are selected as annihilation candidates whose measured TOT values fall between selected range shown by the dashed lines. The selected candidates are further filtered out based on their emission time difference estimated from the centre of the flange. To calculate the emission time of the photon, the hit time of each of the two annihilation candidates is corrected by their estimated TOF. The last criterion applied is based on angular correlation. Hits are marked as back-to-back if the angular difference is between 175$^{0}$ - 180$^{0}$.
 The hit times of tagged annihilation candidates allow the calculation of the TOF as the difference of their arrival time, which is later used to reconstruct the annihilation point on the constructed LOR based on their hit positions. The TOF spectrum obtained is shown in Fig.~\ref{results}. For the current setup, the value of TOF resolution ($\sigma$) is $\approx$ 125~ps. Fig.~\ref{results}(c,d) shows the reconstructed images of the annihilation vertices for the zx and zy planes. The projections on the x, y, and z axes are shown in Fig.~\ref{results}(e). The spatial resolutions ($\sigma$) in the x, y, and z directions are 11, 4, and 13 mm, respectively. Furthermore, since we have access to the $e^{+}$ beam rate, we could estimate the efficiency of two modules in mapping the $e^{+}$annihilation vertices. To calculate the reconstruction efficiency, we first estimated the total number of annihilation pairs incident on the detection modules using the $e^{+}$ beam rate, corrected for the solid angle covered by the two detection units. Finally, the number of entries in the final spectra (Fig.~\ref{results}(c, d)) is divided by the number of total annihilations. The estimated reconstruction efficiency is 14$\%$.

\section{Conclusions and outlook}
We have shown that the modular J-PET with only 2 multi-strip detection units has the potential for imaging the $e^{+}e^{-}$ annihilation spots. The experiment was performed on a $e^{+}$ beamline capable of delivering a continuous monoenergetic $e^{+}$ beam with diameter of a few mm. The obtained TOF and spatial resolution ($\sigma$) are promising for the planned applications on gravitational tests. One of the main objectives of this study was to investigate the ability of the J-PET modules to distinguish annihilation spots that are within 10-20~mm, especially along the beam direction (y-axis). In the present study, we found that the resolution ($\sigma$) along the y-axis is about 4~mm, which is promising for the use of these detection modules as position sensitive detectors for inertial sensing measurements on Ps atoms~\cite{Mariazzi,Sharma}. During these pilot studies, some limitations were identified. To cover the larger solid angles for annihilation photon registration, we placed the modules relatively close to the beamline, which hindered the ability to calibrate the modules with a point source of known activity as strips could not be irradiated uniformly, thus limiting the optimal calibration of the modules. In addition, a dedicated Monte Carlo simulation is required to validate the achieved reconstruction performance of the modules, which can also correctly estimate the attenuation caused by the flanges used and the efficiency of the analysis cuts. Nevertheless, we performed the calibration of the detectors for the first time with cosmic rays and were able to calibrate the detectors. The preliminary results show that the modular multi-strip detectors are able to reconstruct the annihlation vertices of the $e^{+}$ beam spot with an efficiency of 14$\%$, which could be further increased by optimizing the geometry of the detection modules.
%
\section{Acknowledgements}
The authors acknowledge the technical and administrative support of A. Heczko, M. Kajetanowicz and W. Migdał. This work was supported by the Foundation for Polish Science through the TEAM POIR.04.04.00-00-4204/17program, the National Science Centre of Poland through grants MAESTRO no. 2021/42/A/ST2/00423, OPUS no.~2019/35/B/ST2/03562, Miniatura 6 no.2022/06/X/ST2/01444, the Ministry of Education and Science through grant no. SPUB/SP/490528/2021, the EU Horizon 2020 research and innovation programme, STRONG-2020 project, under grant agreement No 824093, and the SciMat and qLife Priority Research Areas budget under the program Excellence Initiative - Research Universityat the Jagiellonian University, and Jagiellonian University project no. CRP/0641.221.2020. B.C.H. acknowledges support of this research by the  Austrian Science Fund (FWF) project P36102-N. The authors also gratefully acknowledge the support of Q@TN, the joint laboratory of the University of Trento, FBK-Fondazione Bruno Kessler, INFN-National Institute of Nuclear Physics, and CNR-National Research Council.

\end{document}